\documentclass[%
 reprint,
 amsmath,amssymb,
 aps,
]{revtex4-1}
\usepackage{psfrag}
\usepackage{color}
\usepackage{graphicx}% Include figure files
\usepackage{dcolumn}% Align table columns on decimal point
\usepackage{bm}% bold math

\begin{document}

\preprint{APS/123-QED}

\title{Space-Time Modulation Induced Non-reciprocity in Electromagnetic Metasurfaces}% Force line breaks with \\
\author{Shulabh Gupta}
 \email{shulabh.gupta@carleton.ca}

\author{Scott. A. Stewart}
\author{Tom J. Smy}%
\affiliation{%
Department of Electronics, Carleton University, 1125 Colonel by Drive, Ottawa, Ontario, Canada\\
}%


\begin{abstract}
Space-time modulation induced non-reciprocity in EM metasurfaces is proposed and numerically demonstrated using rigorous Generalized Sheet Transitions Conditions (GSTCs) under oblique plane-wave incidence. It is phenomenologically shown that the space-time modulation of surface susceptibilities create an inherent asymmetry with respect to the directional perturbation on the metasurface and the transverse wave momentum of the input wave, between forward and backward propagations, resulting in non-reciprocal wave transmission. Exploiting the periodicity of the surface susceptibilities in both time and space, Floquet mode expansion method is used to rigorously compute the scattered fields from, inherently dispersive metasurfaces, by solving GSTCs in combination with causal Lorentzian surface susceptibilities. Various harmonic solutions are shown and the non-reciprocal wave transmission has been confirmed under oblique plane-wave incidence.

\end{abstract}

\pacs{Valid PACS appear here}% PACS, the Physics and Astronomy
\maketitle

\section{Introduction}

Electromagnetic (EM) metasurfaces are two-dimensional equivalents of volumetric metamaterials, and are composed of 2D arrays of sub-wavelength scatterers. By engineering these scatterers across the surface, various interesting wave-shaping transformations can be achieved for various applications such as generalized refraction, holography, polarization control, imaging and cloaking, to name a few \cite{GeneralizedRefraction}\cite{meta3}. Metasurfaces achieve such wave transformations as a result of complex interplay between the electric and magnetic dipolar moments generated by the scatterers, which is sometimes also referred to as a Huygens' configuration \cite{Elliptical_DMS}\cite{Kivshar_Alldielectric}. A convenient implementation of such metasurfaces are using all-dielectric resonators, which naturally produce the electric and magnetic dipoles moments, which when properly designed, provide zero backscattering, resulting in a perfect transmission \cite{Kerker_Scattering}\cite{AllDieelctricMTMS}\cite{Grbic_Metasurfaces}.

A recently growing area of interest is reconfigurable and time-varying metasurfaces, where the constitutive parameters (surface susceptibilities, $\chi_\text{e,m}$) of the metasurfaces are real-time tunable. A more general description of such dynamic metasurfaces, is a space-time modulated metasurface, where the surface susceptibilities are both a function of space and time, resulting in a travelling-type perturbation on the metasurface. They are the 2D equivalents of general space-time modulated mediums \cite{TamirST}\cite{OlinerST}, which have found important applications in parametric amplifiers and acousto-optic spectrum analyzers, \cite{TamirAcoustoDiffraction}\cite{Goodman_Fourier_Optics}\cite{Saleh_Teich_FP}, for instance. Space-time modulation has led to various exotic effects such harmonic generation and non-reciprocity \cite{Caloz_ST_Modulated}\cite{Mechanicl_ST_Modulation}, that has also been recently explored using metasurfaces \cite{STGradMetasurface}\cite{ShaltoutSTMetasurface} for advanced wave-shaping applications. Their attractive features lies in achieving non-reciprocity using purely non-magnetic materials, which has important practical benefits in engineering systems, related to high frequency operation and no requirement of a magnetic bias.

In this work, the space-time modulation induced non-reciprocity is demonstrated and modelled using Generalized Sheet Transition Conditions (GSTCs), which rigorously models the EM behaviour of an ideal zero-thickness metasurface. Since metasurfaces are constructed using sub-wavelength resonators, they are operated around the resonant frequencies where the EM waves have maximum interaction with the metasurface. Consequently, these metasurfaces are naturally very \emph{dispersive}, i.e. $\tilde{\chi}_\text{e,m}(\omega) \ne \text{const.}$ The geometrical shapes of the constituting scatterers are primarily responsible for their resonant behaviour, in spite of their design based on purely non-dispersive materials (typically metals and dielectrics). This operation of the metasurface in a dispersive (and thus broadband) regime, demands a physical description of these resonators consistent with the \emph{causality requirements}. This in turn, requires a causal description of the equivalent surface $\chi_\text{e,m}$ of the metasurface, in frequency (or time domain), i.e. $\tilde{\chi}_\text{e,m} = \tilde{\chi}_\text{e,m}(\omega)$ or $\chi_\text{e,m} = \chi_\text{e,m}(t)$. This requirement is also critical in the accurate time-domain modelling of general space-time modulated metasurfaces, where new spectral frequency components are generated. This subsequently further necessitates a complete description of the surface $\chi_\text{e,m}$ encompassing these frequencies as well, in addition to the bandwidth of the input excitation. 

In this context, the non-reciprocal behaviour of a general space-time modulated metasurface is demonstrated here assuming Lorentzian surface susceptibilities, which are naturally causal and rigorously capture the fundamentally dispersive nature of EM metasurfaces. This work presents space-time modulation induced non-reciprocity in EM metasurfaces from a phenomenological point of view, and demonstrating the predictions using rigorous characterization of scattering fields produced by physically motivated dispersive metasurfaces.

\section{Space-Time Modulation induced Non-reciprocity}

\subsection{Metasurface Description}

\begin{figure}[htbp]
\begin{center}
\psfrag{a}[c][c][0.8]{$z=0$}
\psfrag{z}[c][c][0.8]{$z$}
\psfrag{x}[c][c][0.8]{$x$}
\psfrag{t}[c][c][0.8]{$t$}
\psfrag{b}[c][c][0.8]{$\psi(\mathbf{r}, t) = \psi_0(\mathbf{r}, t) \sin(\omega_0t)$}
\psfrag{f}[c][c][0.6]{$\boxed{\chi_{ee}(x, t),\; \chi_{mm}(x, t)}$}
\psfrag{d}[c][c][0.6]{$\psi_2(\mathbf{r}, t) \sin\{(\omega_0 - \omega_p)t\}$}
\psfrag{c}[l][c][0.6]{$\psi_1(\mathbf{r}, t) \sin\{\omega_0 t\}$}
\psfrag{e}[c][c][0.6]{$\psi_3(\mathbf{r}, t) \sin\{(\omega_0 + \omega_p)t\}$}
\includegraphics[width=0.65\columnwidth]{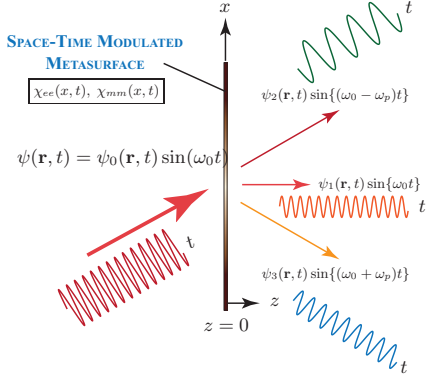}
\caption{A general Illustration of a space-time modulated metasurface under oblique plane-wave incidence resulting in generation of several frequency harmonics, refracted along different angles. $\omega_p$ is the pumping frequency.}\label{Fig:Problem}
\end{center}
\end{figure}

Consider a metasurface of Fig.~\ref{Fig:Problem}, which is excited with an incident plane-wave at an oblique angle for a given excitation frequency $\omega_0$. The metasurface can be described using electric and magnetic surface susceptibilities, $\chi_\text{ee}(x, t),\; \chi_\text{mm}(x, t)$, which both are periodic functions of space and time, for a general space-time modulated metasurface. Such a metasurface is known to produce several harmonic frequencies in the scattered fields refracted along different directions, as illustrated in Fig.~\ref{Fig:Problem}, and demonstrated in \cite{Stewart_Metasurface_STM}\cite{Gupta_Harmonic_MS}.

Metasurfaces can be seen as space-discontinuities, and their exact EM characteristics can be rigorously modelled using GSTCs \cite{IdemenDiscont}\cite{KuesterGSTC}. For the metasurface configuration illustrated in Fig.~\ref{Fig:Problem}, the time-domain GSTCs are given by \cite{MS_Synthesis}

\begin{subequations}\label{Eq:GSTC}
\begin{equation}\label{Eq:GSTC_1}
\hat{\mathbf{z}}\times\Delta \mathbf{H} = \epsilon_0\frac{d\mathbf{Q}_{||}}{dt} - \hat{\mathbf{z}} \times \nabla_{||}M_z
\end{equation}
\begin{equation}\label{Eq:GSTC_2}
\Delta \mathbf{E} \times \hat{\mathbf{z}} = \mu_0\frac{d\mathbf{M}_{||}}{dt} - \nabla_{||}Q_z \times \hat{\mathbf{z}},
\end{equation}
\end{subequations}

\noindent where $\Delta\psi$ and $\psi_\text{av}$ are the field difference and average, between the total fields in transmission and reflection, respectively; $\mathbf{Q}_{||}(t) = \mathcal{F}^{-1}\{\tilde{\mathbf{Q}}_{||}(\omega)=\tilde{\chi}_\text{ee}(\omega)\tilde{\mathbf{E}}_\text{av}(\omega)\}$ and $\mathbf{M}_{||}(t) = \mathcal{F}^{-1}\{\tilde{\mathbf{M}}_{||}(\omega)=\tilde{\chi}_\text{mm}(\omega)\tilde{\mathbf{H}}_\text{av}(\omega)\}$ are the tangential electric (normalized) and magnetic surface polarizations, in terms of surface susceptibilities; and $Q_z$ and $M_z$ are the longitudinal polarization densities. For simplicity in the subsequent analysis we make an assumption that the longitudinal polarizations are zero, i.e. $P_z = M_z = 0$. This assumption has no fundamental implications on the results of this work.

\subsection{Principle}

A space-time modulated metasurface can be described by introducing a travelling periodic modulation of the surface susceptibilities such that

\begin{align}
\chi_\text{ee}(x, t) &= \chi_\text{ee}\left\{\omega_m t - \beta_m x\right\}\notag\\
\chi_\text{mm}(x, t) &= \chi_\text{mm}\left\{\omega_m t - \beta_m x\right\},\label{Eq:STMod}
\end{align}

\noindent where $\omega_m$ and $\beta_m$ are the modulation (or pumping) frequency and modulation wavenumber, respectively. Here a 2D problem is assumed throughout for simplicity, where any variation of the field quantities is zero along the $y-$direction. A good choice of a causal and a physical description of static time-domain surface susceptibilities is a Lorentz response given by

\begin{subequations}\label{Eq:Lorentz}
\begin{equation}
\tilde{\chi}_\text{ee}(\omega)  = \frac{\omega_p^2}{(\omega_r^2 - \omega^2) + j\alpha\omega} = \frac{\tilde{Q}(\omega)}{\tilde{E}(\omega)},~\text{or}
\end{equation}
\begin{equation}
\frac{d^2Q(t)}{dt^2} +\alpha \frac{dQ(t)}{dt} + \omega_r^2 Q(t) = \omega_p^2 E(t) \\
\end{equation}
\end{subequations}

\noindent in either frequency or time domain, where $\omega_p$, $\omega_r$ and $\alpha$ are the plasma frequency, resonant frequency and the loss coefficient, respectively. While the susceptibilities are modelled here using a single Lorentzian contribution for simplicity, a more practical dispersion profile can easily be constructed using multiple Lorentzians. The space-time modulation of the surface susceptibilities can now  be introduced by varying the parameters of these equivalent Lorentz resonators. In this work for simplicity, we assume that its the resonant frequency only, that is space-time varied following a periodic modulation given by

\begin{equation}
\omega_r(x,t) = \omega_r\{1 + \Delta_m\cos(\omega_mt - \beta_m x)\},
\end{equation}

\noindent where $\Delta_m$ is the modulation index with $|\Delta_m|\le 1$, and $\omega_r$ is the static value of the resonant frequency.

\begin{figure}[htbp]
\begin{center}
\psfrag{L}[c][c][0.6]{$K_x^\text{BWD}$}
\psfrag{B}[c][c][0.6]{$K_z^\text{FWD}$}
\psfrag{A}[c][c][0.6]{$K_z^\text{BWD}$}
\psfrag{D}[c][c][0.75]{$E_0$}
\psfrag{F}[c][c][0.75]{$E_t$}
\psfrag{E}[c][c][0.75]{$H_t$}
\psfrag{x}[c][c][0.75]{$x$}
\psfrag{z}[c][c][0.75]{$z$}
\psfrag{J}[c][c][0.6]{$K_x^\text{FWD}$}
\psfrag{a}[c][c][0.75]{$\mathbf{m_x}~[\chi_{mm}]$}
\psfrag{b}[c][c][0.75]{$\mathbf{p_y}~[\chi_{ee}]$}
\psfrag{c}[c][c][0.75]{$\mathbf{k_z}$}
\psfrag{H}[c][c][0.75]{$\mu_0,~\epsilon_0$}
\psfrag{K}[r][c][0.75]{$\chi(x,t)$}
\includegraphics[width=0.6\columnwidth]{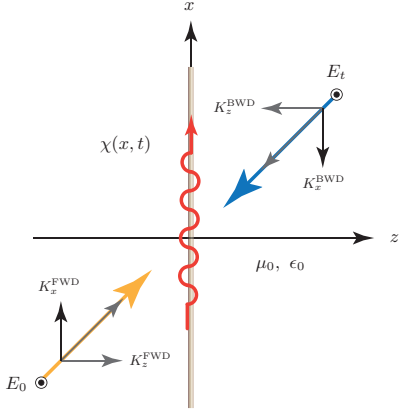}
\caption{Intuitive explanation of non-reciprocal behaviour in space-time modulated metasurfaces. The transverse momentum of the incident wave, $K_x^\text{FWD}$, is parallel to the direction of surface susceptibility perturbation (along $+x$) in the forward transmission, compared to being anti-parallel in the backward transmission. Metasurface is assumed to be matched.}\label{Fig:NR}
\end{center}
\end{figure}

Let us consider the illustration of Fig.~\ref{Fig:NR}, where an oblique propagating plane-wave at $\omega_0$, is incident on a \emph{matched} space-time modulated metasurface. For a static metasurface, according to Snell's law, $\sin\theta_\text{out} = \sin\theta_\text{in}$, so that the transmitted wave has the same angle of propagation as the input. Now lets introduce the space-time modulation on the metasurface. Due to the fixed and specified direction of the susceptibility perturbation on the metasurface, the transverse wave momentum in forward transmission, $K_x^\text{FWD}$, is \emph{parallel} to the susceptibility perturbation along $+x$. However, when the same wave is launched back from the transmission side at the same incidence angle, the transverse wave momentum, $K_x^\text{BWD}$, is \emph{anti-parallel} to the susceptibility perturbation. Therefore the scattered fields in the forward transmission mode are expected to be different from the fields in the backward transmission mode. This implies that \emph{the space-time modulated metasurface is inherently non-reciprocal under oblique incidence}. The symmetry is naturally restored for the case of normal incidence, where non-reciprocity cease to exist.

\subsection{Field Equations}

\begin{figure}[htbp]
\begin{center}
\psfrag{A}[c][c][1]{$H_0$}
\psfrag{B}[r][c][1]{$H_r$}
\psfrag{D}[c][c][1]{$E_0$}
\psfrag{F}[c][c][1]{$E_{t,n}$}
\psfrag{E}[c][c][1]{$H_{t,n}$}
\psfrag{G}[c][c][1]{$E_r$}
\psfrag{x}[c][c][1]{$x$}
\psfrag{z}[c][c][1]{$z$}
\psfrag{J}[c][c][1]{$\theta_1$}
\psfrag{a}[c][c][1]{$\theta_0$}
\psfrag{b}[c][c][1]{$\mathbf{p_y}~[\chi_{ee}]$}
\psfrag{C}[r][c][1]{$\chi = f(x-vt)$}
\psfrag{I}[c][c][1]{$n^\text{th}$-harmonic}
\psfrag{H}[c][c][1]{$\mu_0,~\epsilon_0$}
\psfrag{K}[c][c][1]{$\theta_n$}
\includegraphics[width=0.75\columnwidth]{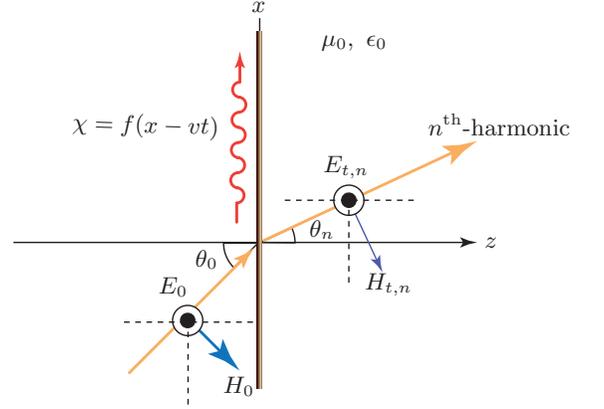}
\caption{Field configuration showing the $n^\text{th}$ harmonic of the Floquet expansion of \eqref{Eq:FloquetEt}, for an oblique plane-wave interacting with a space-time modulated metasurface.}\label{Fig:ObliqueTE}
\end{center}
\end{figure}

To rigorously demonstrate the non-reciprocal nature of space-time modulated metasurfaces, consider a Transverse Electric (TE) plane-wave incident on an infinite metasurface as shown in Fig.~\ref{Fig:ObliqueTE}. The incident tangential fields are given (time convention here is $e^{j\omega t}$) by

\begin{align}\label{Eq:E0}
\mathbf{E}_0(x,z=0_-, \omega) &= E_0e^{-jk_0\sin\theta_0x}  ~\mathbf{\hat{y}},\notag\\
\mathbf{H}_{0, ||}(x,z=0_-, \omega) &= -\frac{E_0\cos\theta_0}{\eta_0}e^{-jk_0\sin\theta_0x}  ~\mathbf{\hat{x}}
\end{align}

\noindent For the assumed matched metasurface with $R=0$, this periodic modulation of the susceptibility implies that the transmitted fields at $z=0_+$ will also be periodic in both space and time so that they can be expanded in terms of a Floquet series as \cite{Rothwell},

\begin{align}
\mathbf{E}_t(x, 0_+, t) = \sum_{n=-\infty}^{\infty} b_ne^{j(\omega + n\omega_m)t}e^{-j (k_0\sin\theta_0 + n\beta_m) x}~\mathbf{\hat{y}},\label{Eq:FloquetEt}
\end{align}

\noindent  Each term of this expansion represents a plane-wave at a frequency $\omega_n = (\omega + n\omega_m)$ travelling at an oblique angle $\theta_n$ given by

\begin{align}
\sin\theta_n =  \left(\frac{k_0\sin\theta_0 + n\beta_m}{k_0 + nk_m}\right). 
\end{align}

\noindent  For propagating field solutions in the transmission half-space, $-1 \le \sin\theta_n \le 1$, so that the corresponding ranges of harmonic index $n$ are given by

\begin{align}
n_\text{min} = -\frac{k_0(1 + \sin\theta_0)}{\beta_m + k_m} \le n \le \frac{k_0(1 - \sin\theta_0)}{\beta_m - k_m} = n_\text{max}.\label{Eq:FWDn}
\end{align}

\noindent Rest of the Floquet components correspond to surface waves on the metasurface and evanescent along the longitudinal $z-$direction. Finally, the \emph{tangential} H-fields corresponding to the transmitted scattered E-fields of \eqref{Eq:FloquetEt}, can be written as

\begin{align}
\mathbf{H}_{t,||}(x,t) = -\sum_{-\infty}^{\infty} \frac{E_n}{\eta_0}\cos\theta_n~\mathbf{\hat{x}}\label{Eq:FloquetHt}\notag\\
\text{with}~\cos\theta_n = \sqrt{1 - \left(\frac{k_0\sin\theta_0 + n\beta_m}{k_0 + nk_m}\right)^2}.
\end{align}

\noindent Substituting \eqref{Eq:E0}, \eqref{Eq:FloquetEt} and \eqref{Eq:FloquetHt} in \eqref{Eq:GSTC}, we get

\begin{align}
& \frac{\epsilon_0}{2}\left\{\frac{dQ_{t, ||}(x,t)}{dt} +\frac{dQ_0(x,t)}{dt}\right\} = H_{t,||}(x,t) + \frac{E_0(x,t)\cos\theta_0}{\eta_0},\label{Eq:FE_1}
\end{align}

\noindent which is the first field equation, where $\mathbf{Q}_{t, ||}(x,t) = \mathcal{F}^{-1}\{\tilde{\mathbf{Q}}_{t, ||}(x, \omega)=\tilde{\chi}_\text{ee}(\omega)\tilde{\mathbf{E}}_t(x,\omega)\}$ and $\mathbf{Q}_{0, ||}(x,t) = \mathcal{F}^{-1}\{\tilde{\mathbf{Q}}_{0, ||}(x, \omega)=\tilde{\chi}_\text{ee}(\omega)\tilde{\mathbf{E}}_0(x,\omega)\}$, respectively. It should be noted that while we have used the first choice of the GSTC equation \eqref{Eq:GSTC_1} here, the second choice of \eqref{Eq:GSTC_2} could have also been used in terms of magnetic susceptibility, which also follows a Lorentzian response. In that case of second choice, the magnetic susceptibility must be carefully defined in relation to the specified electric susceptibility, in order to maintain the underlying assumption of a matched metasurface. For the general case of a mismatched metasurface, both GSTCs must be considered to include the non-zero reflection.

Following the assumed Lorentzian susceptibilities of \eqref{Eq:Lorentz}, we get 

\begin{align}
&\frac{d^2Q_{0/t, ||}(x,t)}{dt^2} +\omega_r^2(x,t)Q_{0/t, ||}(x,t) = \omega_p^2 E_{0,t}(x,t).\label{Eq:FE_2}
\end{align}

\noindent for each of the transmitted and incident surface polarizabilities. Equations~\eqref{Eq:FE_1} and \eqref{Eq:FE_2} represents in total three field equations, for the primary unknown $E_t(x,t)$, and two auxiliary unknowns $Q_{0, ||}(x,t)$ and $Q_{t, ||}(x,t)$. It should be noted, that the assumption of a matched metasurface has led to this simple set of field equations, either in terms of purely electric or magnetic surface polarizabilities.

\begin{figure}[htbp]
\begin{center}
\psfrag{A}[l][c][0.9]{$\chi = f(x+vt)$}
\psfrag{B}[r][c][0.9]{$H_r$}
\psfrag{D}[c][c][0.9]{$E_0^+(\omega_0)$}
\psfrag{F}[c][c][0.9]{$E_0^-(\omega_0)$}
\psfrag{E}[c][c][0.9]{$H_{t,n}$}
\psfrag{G}[c][c][0.9]{$E_r$}
\psfrag{x}[c][c][0.9]{$x$}
\psfrag{z}[c][c][0.9]{$z$}
\psfrag{J}[c][c][0.9]{$\theta_1$}
\psfrag{a}[c][c][0.9]{$\theta_0$}
\psfrag{b}[c][c][0.9]{$\mathbf{p_y}~[\chi_{ee}]$}
\psfrag{C}[r][c][0.9]{$\chi = f(x-vt)$}
\psfrag{I}[c][c][0.9]{$n^\text{th}$-harmonic}
\psfrag{H}[c][c][0.9]{$\mu_0,~\epsilon_0$}
\psfrag{K}[c][c][0.9]{$\theta_n$}
\includegraphics[width=0.6\columnwidth]{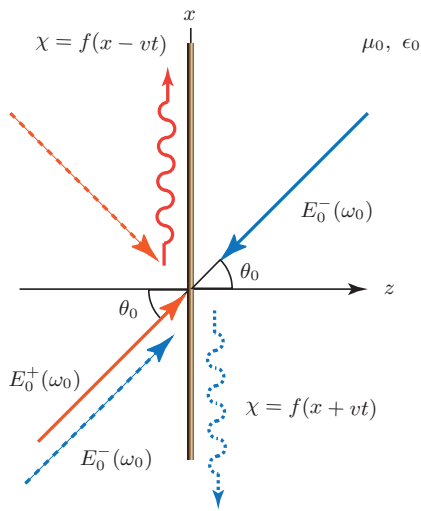}
\caption{Field configuration for testing the non-reciprocal characteristics of a space-time modulated metasurface. Solid orange curve corresponds to forward transmission, and the dashed blue curve corresponds to an equivalent backward transmission, i.e. $\beta_m \rightarrow -\beta_m$.}\label{Fig:Checks}
\end{center}
\end{figure}

The non-reciprocal nature of the space-time modulated metasurfaces can be inferred also from direct observation of these field equations, as follows. The above field equations are developed naturally for forward transmission with $E_0(\omega_0)^+$ as shown in Fig.~\ref{Fig:Checks}. The backward transmission with $E_0(\omega_0)^-$, through the metasurface, on the other hand, can be equivalently modelled in terms of forward transmission by reversing the direction of the susceptibility modulation, i.e. $\beta_m\rightarrow -\beta_m$ for same $\theta_0$. This results in: 1) a different range of the harmonic indices $n$ in the Floquet series of \eqref{Eq:FloquetEt}, given by

\begin{align}
-\frac{k_0(1 - \sin\theta_0)}{\beta_m + k_m} \le n \le \frac{k_0(1 + \sin\theta_0)}{\beta_m - k_m},
\end{align}

\noindent compared to that for forward transmission, i.e. \eqref{Eq:FWDn}. 2) The first field equation \eqref{Eq:FE_1}, is modified through $H_{t,||}(x,t)$ as the definition of $\cos\theta_n$ changes to,

\begin{align*}
\cos\theta_n = \sqrt{1 - \left(\frac{k_0\sin\theta_0 - n\beta_m}{k_0 + nk_m}\right)^2} %\ne \eqref{Eq:FloquetHt}.
\end{align*}

\noindent modifying \eqref{Eq:FloquetHt}. Therefore, the forward and backward wave transmissions are governed by two different set of field equations, from which different scattered fields are naturally expected. It should also be noted from Fig.~\ref{Fig:Checks}, that the backward transmission is also equivalent to maintaining the direction of modulation along $+x$, but reversing the angle of incidence, i.e. $\theta_0 \rightarrow -\theta_0$ for same $\beta_m$. It can be easily verified that the resulting ranges of indices $n$ and the field equations are identical to when $\beta_m\rightarrow -\beta_m$ for same $\theta_0$.

\begin{figure}[htbp]
\begin{center}
\psfrag{a}[c][c][0.75]{harmonic index, $n$}
\psfrag{b}[c][c][0.75]{Floquet coefficient, $|b_n|$}
\psfrag{c}[l][c][0.65]{$\beta_m > 0$}
\psfrag{d}[l][c][0.65]{$\beta_m < 0$}
\psfrag{e}[c][c][0.75]{Normal incidence, $\theta_0=0^\circ$}
\psfrag{f}[c][c][0.75]{$+x$ modulation, $\beta_m > 0$}
\psfrag{m}[l][c][0.65]{$\theta_0=0^\circ$}
\psfrag{n}[l][c][0.65]{$\theta_0 = 45^\circ$}
\includegraphics[width=\columnwidth]{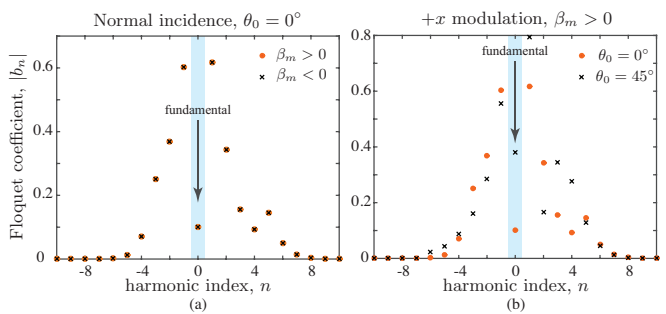}
\caption{Output harmonic spectrum in forward transmission, computed using the three fields equations of \eqref{Eq:Field_Set} for the case of plane-wave a) with normal incidence for modulation along $+x$ and $-x$, respectively. b) with $+x$ modulation $\beta_m>0$, for two angles of incidence, $\theta_0 = 0^\circ$ and $\theta_0=45^\circ$. The Lorentzian parameters are: $\omega_p = 2\pi (50~\text{GHz})$, $\omega_{r} = 2\pi(225~\text{THz})$, the excitation frequency $\omega_0=2\pi(230~\text{THz})$, the pumping frequency $\omega_m = 0.025\omega_0$ and the pumping wave number $\beta_m = 10\pi/(25\mu\text{m})$. Only first $\pm10$ harmonics are shown along with the fundamental. }\label{Fig:Normal}
\end{center}
\end{figure}

\begin{figure}[htbp]
\begin{center}
\psfrag{a}[c][c][0.75]{harmonic index, $n$}
\psfrag{b}[c][c][0.75]{Floquet coefficient, $|b_n|$}
\psfrag{c}[l][c][0.65]{$\beta_m > 0$}
\psfrag{d}[l][c][0.65]{$\beta_m < 0$}
\psfrag{e}[c][c][0.75]{$\theta_0=30^\circ$}
\psfrag{f}[c][c][0.75]{$\theta_0=45^\circ$}
\psfrag{m}[l][c][0.65]{$\Delta b_0$}
\psfrag{n}[l][c][0.65]{$\theta_0 = 45^\circ$}
\includegraphics[width=\columnwidth]{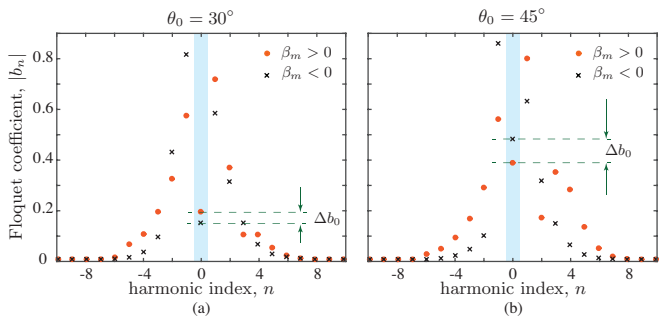}
\caption{Output harmonic spectrum in forward and equivalent backward transmission for an obliquely incident plane-wave at an angle of a) $\theta_0 = 30^\circ$ and b) $\theta_0 = 45^\circ$. The metasurface parameters are the same as in Fig.~\ref{Fig:Normal}. Both propagating and evanescent harmonics are shown}\label{Fig:Result}
\end{center}
\end{figure}

\section{Numerical Demonstration}

\subsection{Field Solutions}

To solve the field equations derived in the previous section, let us expand the surface polarizabilities corresponding to incident and transmitted fields using Floquet series as,

\begin{subequations}\label{Eq:P0PtFl}
\begin{equation}
Q_0(x,  t) = \sum_{-\infty}^{\infty} a_n e^{j (\omega + n\omega_m) t}e^{-j(k_0\sin\theta + n\beta_m) x}
\end{equation}
\begin{equation}
Q_t(x,   t) = \sum_{-\infty}^{\infty} c_n e^{j (\omega + n\omega_m) t}e^{-j(k_0\sin\theta + n\beta_m) x},
\end{equation}
\end{subequations}

\noindent where $a_n$ and $b_n$ are unknown coefficients to be determined. Substituting the Floquet forms of all the three unknowns, \eqref{Eq:FloquetEt} and \eqref{Eq:P0PtFl}, in the field equations \eqref{Eq:FE_1} and \eqref{Eq:FE_2}, and rearranging the terms with $|E_0|=1$, we get \eqref{Eq:Field_Set_1}, \eqref{Eq:Field_Set_2} and \eqref{Eq:Field_Set_3}, where the series is truncated to $(2N+1)$ terms. These three equations represent three set of linear equations. Since the $n^\text{th}$ coefficient in \eqref{Eq:Field_Set_1}, \eqref{Eq:Field_Set_2} corresponding to Lorentzian relations, are coupled to neighbouring coefficients, $(n = \pm 1)$ and $(n = \pm 2)$, the following boundary conditions are enforced to solve the required unknowns:

\begin{figure*}
\begin{subequations}\label{Eq:Field_Set}
\begin{equation}\label{Eq:Field_Set_1}
\sum_{-N}^{N} \left[\frac{\Delta_m^2\omega_r^2}{4\omega_p^2}a_{n-2} +  \frac{\Delta_m\omega_r^2}{\omega_p^2} a_{n-1} +  \left\{\frac{\omega_r^2}{\omega_p^2} - \frac{(\omega_0 + n\omega_m)^2}{\omega_p^2} +  \frac{\Delta_m^2\omega_r^2}{2\omega_p^2}\right\}a_n + \frac{\Delta_m\omega_r^2}{\omega_p^2} a_{n+1} + \frac{\Delta_m^2\omega_r^2}{4\omega_p^2}a_{n+2} \right]e^{jn\omega_mt}e^{-jn\beta_mx}  = 1
\end{equation}
\begin{equation}\label{Eq:Field_Set_2}
\sum_{-N}^{N} \left[\frac{\Delta_m^2\omega_r^2}{4\omega_p^2}c_{n-2} +  \frac{\Delta_m\omega_r^2}{\omega_p^2} c_{n-1} +  \left\{ \left(\frac{\omega_r^2}{\omega_p^2} - \frac{(\omega_0 + n\omega_m)^2}{\omega_p^2} +  \frac{\Delta_m^2\omega_r^2}{2\omega_p^2} \right)c_n -  b_n\right\} + \frac{\Delta_m\omega_r^2}{\omega_p^2} c_{n+1} + \frac{\Delta_m^2\omega_r^2}{4\omega_p^2}c_{n+2} \right]e^{jn\omega_mt} e^{-jn\beta_mx} = 0
\end{equation}
\begin{equation}\label{Eq:Field_Set_3}
\sum_{-N}^{N} \left\{j\frac{(\omega_0 + n\omega_m)}{2c}a_n + \sqrt{1 - \left(\frac{k_0\sin\theta_0 + n\beta_m}{k_0 + nk_m}\right)^2} b_n + j\frac{(\omega_0 + n\omega_m)}{2c} c_n\right\}e^{j n\omega_mt}e^{-jn\beta_mx} = \cos\theta_0
\end{equation}
\end{subequations}
\end{figure*}

\begin{align}
&a_{-N-1}=a_{-N-2} =0 ,~
&a_{N+1} = a_{N+2} = 0.
\end{align} 

\noindent These boundary conditions make the assumption that N is sufficiently large that coupling into these modes is negligible. With these conditions, the three sets of equations can now be easily expressed in terms of a matrix form and solved for the unknown Floquet coefficients.

\subsection{Results}

Let us consider an initial case of normal plane-wave incidence on a non-modulated metasurface. Substituting $\Delta_m=0$ in \eqref{Eq:Field_Set_1}, \eqref{Eq:Field_Set_2} and \eqref{Eq:Field_Set_3}, and solving for $n=0$ harmonic, gives

\begin{equation}
b_0(\omega_0) = \left(\frac{2c(\omega_r^2 - \omega_0^2) - j\omega_0\omega_p^2}{2c(\omega_r^2 - \omega_0^2) + j\omega_0\omega_p^2}\right).
\end{equation}

\noindent This is an all-pass (phase-only) transfer function with $|b_0|=1$, as expected. Next, let us add the modulation, while maintaining the normal plane-wave incidence (with zero transverse momentum). From the illustration of Fig.~\ref{Fig:Checks}, it can be expected that the direction of modulation should have no impact on the strength of the scattered fields, in this case. Fig.~\ref{Fig:Normal}(a) confirms this prediction and shows the exact same strength of the Floquet harmonics for susceptibility modulation along either $+x$ or $-x$ direction. However, the situation is different, between normal ($k_x =0$) and oblique plane-wave incidence ($k_x\ne 0$), so that the harmonic strengths of the scattered fields vary greatly with specified angle of incidences, as shown in Fig.~\ref{Fig:Normal}(b).

Next, to test the non-reciprocal behaviour of the metasurface, As explained in Sec. II-B, the field equations are solved for two situations modelling a) forward transmission with $\theta_0\ne0$ and $\beta_m >0$, and b) an equivalent backward transmission with $\theta_0\ne0$ and $\beta_m < 0$. Fig.~\ref{Fig:Result} shows the computed results for two cases of angle of incidences. As expected, for the given choice of modulation parameters, the harmonic strengths are different between forward and backward transmissions ($\Delta b_0\ne 0$), confirming the non-reciprocity in space-time modulated metasurfaces. In more practical terms, if a wave at $\omega_0$ is incident on a metasurface at an angle of $\theta_0$, its transmission coefficient is different from the case, when it is transmitted back from the opposite side of the surface at the same angle and frequency. It should also be noted that, apart from the fundamental frequency of interest, all other harmonics also have significantly different strengths between forward and backward transmission.

\section{Conclusions}

Space-time modulation induced non-reciprocity in EM metasurfaces has been proposed and numerically demonstrated using rigorous GSTCs for oblique plane-wave incidence. It has been phenomenologically shown that the space-time modulation of surface susceptibilities create an inherent asymmetry with respect to the directional perturbation on the metasurface and the transverse wave momentum of the input wave, between forward and backward propagations, resulting in non-reciprocal wave transmission. Exploiting the periodicity of the surface susceptibilities in both time and space, Floquet mode expansion method has been used to rigorously compute the scattered fields from the metasurface by solving GSTCs in combination with Lorentzian surface susceptibilities. Various harmonic solutions have been shown and the non-reciprocal wave transmission has been confirmed under oblique plane-wave incidence.

No effort has been made in this work to optimize or enhance the non-reciprocal wave transmission through the metasurface at the fundamental frequency, by varying the modulation parameters $\omega_m$ and $\beta_m$ for a given choice of Lorentzian parameters of the surface susceptibility. Given the complexity of the problem and large number of physical parameters involved, more extensive work on maximizing and exploiting non-reciprocity from space-time modulated metasurfaces, is required. However, the Floquet mode expansion method presented here-in, provides an ideal analytical tool to investigate the potential applications of space-time modulation in various wave-shaping applications. 

%\section*{Appendix}
%
%\begin{align}
%& \frac{\epsilon_0}{2}\left\{\frac{dQ_{t, ||}(x,t)}{dt} +\frac{dQ_0(x,t)}{dt}\right\} = H_{t,||}(x,t) + \frac{E_0(x,t)\cos\theta_0}{\eta_0}, 
%\end{align}
%
%\begin{align}
%&\frac{d^2Q_{0/t, ||}(x,t)}{dt^2} +\omega_r^2(x,t)Q_{0/t, ||}(x,t) = \omega_p^2 E_{0,t}(x,t). 
%\end{align}
%
%\begin{align}
%& \frac{\mu_0}{2}\left\{\frac{dM_{t, ||}(x,t)}{dt} +\frac{dM_0(x,t)}{dt}\right\} = E_{t,||}(x,t) - E_0(x,t),
%\end{align}
%
%\begin{align}
%&\frac{d^2M_{0/t, ||}(x,t)}{dt^2} +\omega_{m,r}^2(x,t)M_{0/t, ||}(x,t) = \omega_{mp}^2 H_{0/t, ||}(x,t).\label{Eq:FE_2}
%\end{align}

\bibliography{Gupta_ObliqueMetasurface_PRB_2017}% Produces the bibliography via BibTeX.

\end{document}